# The gravitational waves are fictitious entities - III

A. LOINGER

Dipartimento di Fisica, Università di Milano

Via Celoria, 16, 20133 Milano, Italy

ABSTRACT. — Only Levi-Civita's electromagnetic interpretation of the characteristic hypersurfaces of Einstein field equations is conceptually correct.

**1.** – The functions $z(x)$, $[x \equiv (x^0, x^1, x^2, x^3)]$, of the characteristic hypersurfaces $z(x)=0$ of Einstein field equations are solutions of the Hamiltonian equation

$$(1.1) \qquad H := \tfrac{1}{2}\, g^{jk}(x)\, \frac{\partial z(x)}{\partial x^j}\, \frac{\partial z(x)}{\partial x^k} = 0 \;.$$

When $g^{jk}(x)$ has a non-undulatory form the only logical interpretation of $z(x)=0$ is that of Levi-Civita [1]: $z(x)=0$ gives the law of motion of an *electromagnetic* wave front (or of the wave front of any field capable of transmitting signals, *different from the gravitational field*). This is quite obvious, e.g., for the $ds^2$ of Schwarzschild's space-time, for which we have functions $z(x)$ given – in the usual Droste-Weyl co-ordinates – by the expressions (see [2])

$$ct \;\pm\; r - r_* + 2\alpha\, \ln \frac{r - 2\alpha}{r_* - 2\alpha}, \qquad (r > 2\alpha),$$

where $\alpha := GM/c^2$, and $r_* > 2\alpha$ is an integration constant.

When $g^{jk}(x)$ has a wavy form, I do not see any reason to repudiate the above interpretation, because *the undulatory character depends on the reference frame*.

Remark that Levi-Civita's conception is *the* reasonable extension of that valid for the null lines of *special* relativity. Thus also general relativity contains the fundamental law of geometric optics, *and independently of Maxwell equations*. On the contrary, it does not contain the law of a gravitational "geometric optics", as it was shown in [2]. And *a fortiori* the laws of a gravitational "undulatory optics".

**2.** – It is obvious that we can confer some wavy form on any non-undulatory metric tensor – in particular, on a *static* tensor – by an appropriate change of reference frame. And *vice-versa*, it is always possible to reduce a wavy $g^{jk}(x)$ to a non-undulating form through a suitable change of general co-ordinates. Observe that also from the physical point of view the original frame does not possess any privilege.

**3.** – Nowadays the overwhelming majority of the theoreticians are convinced of the limited value of the linearized version of Einstein field equations. However, many experimenters are still very fond of its simple formulae. I wish now to emphasize a straightforward argument, which evidences a basic drawback of the linearized theory. The starting point of this theory is the following, as it is well known: we put

(3.1) $$g_{jk}(x) \approx \eta_{jk} + \varepsilon\, h_{jk}(x) \;,$$

where $\eta_{jk}$ is the usual diagonal Minkowski tensor and $\varepsilon$ is a small parameter. It follows that the curvature tensor $R_{mnrs}$ is approximately given by

(3.2) $$2\, R_{mnrs} \approx \varepsilon\, (\, h_{ms,nr} - h_{ns,mr} - h_{mr,ns} + h_{nr,ms}\,)\;.$$

Now, the approximation (3.1) has a tensorial character only with respect to *Lorentz* transformations and to *infinitesimal* transformations of co-ordinates. Thus, if we perform a change of reference frame $x \rightarrow x'$ not subject to these restrictions, the curvature tensor is no longer given by the simplified expressions (3.2), because the additional terms $2\Gamma^j_{jms} \Gamma^j_{nr} - 2\Gamma_{jmr} \Gamma^j_{ns}$ cannot in any case be neglected. A significant consequence: the plane gravitational waves of the linearized theory have a curvature tensor different from zero, but this is not a meaningful result because this theory is *not* covariant under *all* the co-ordinate changes.

**4.** − Many theoreticians have finally realized the truth of an Eddington's sentence, according to which the pseudo energy tensor of the gravitational field is only a mathematical fiction. Consequently, they strive to excogitate tricks for the detection of the gravitational waves based on the effects produced by a curvature tensor different from zero. But they neglect two facts: *i*) the undulatory structure of a curvature tensor depends on the frame − typically, the harmonic frame; *ii*) the energy balance (of the *real*, not pseudo, energy) in any ideal detection experiment of the gravitational waves is a riddle wrapped up in a mystery. (And likewise "mysterious" is the generation mechanism of the gravitational radiation − the standard acceleration mechanism can be justified only in the linearized version of the theory).

*Only Levi-Civita's electromagnetic interpretation of eq. (1.1) is conceptually correct.*

**APPENDIX A**

In regard to the generation mechanism of the hypothetic gravitational waves, a result of a paper by Bonnor and Swaminarayan [3] is rather interesting.

These authors discovered an exact solution of Einstein field equations *in vacuo*, referring to four mass points moving with different uniform accelerations relative to a frame which is Minkowskian at infinity, except in certain directions. Their analysis "demands the acceptance of a total [gravitational] potential (1/2) (advanced+retarded)." Thus "the system as a whole does not lose energy by radiation". This indicates that − from the standpoint of the exact theory − the acceleration mechanism for the production of gravitational waves is a mere fib.

A discussion on the above paper [3] with Dr S. Antoci is gratefully acknowledged.

**APPENDIX B**

As it is well known, Lorentz [4] and Levi-Civita [5] proposed − and with excellent mathematical reasons − to *define* as energy-momentum of the gravitational field the tensor $(1/\kappa) G_{jk}$, where $G_{jk} := R_{jk} − (1/2) g_{jk} R$, so that Einstein field equations

$$(\text{B.1}) \qquad T_{jk} + \frac{1}{\kappa} G_{jk} = 0 \quad , \qquad (\kappa := 8\pi G/c^2) \quad ,$$

would express simply the conservation of the *total* energy-momentum tensor.

Einstein objected that in such a way the energy-momentum of a closed system would always be equal to zero, and this fact would not imply the further existence of the system under whatever form [6]. However, from the standpoint of the *coherence* of the mathematical formalism, Lorentz and Levi-Civita were undoubtedly right. It is sufficient to recall that in any action principle of any physical theory referred to general co-ordinates, the coefficients of the variations $\delta g^{jk}$ of the components of the fundamental tensor are the components, say $E_{jk}$, of the energy-momentum tensor of the considered field. In general relativity this property is just possessed, *in vacuo*, by the tensor $G_{jk}$.

According to the conception of Lorentz and Levi-Civita, Einstein field equations in empty space (without singularities)

(B.2) $$G_{jk} = 0$$

would admit of no non-trivial *physical* solution, in particular no gravitational radiation could exist.

**APPENDIX C**

In a previous paper [7] I have pointed out that if, e.g., the terrestrial rotation generated the emission of gravitational waves, one would conclude that *spatium est absolutum*, contrary to the fundamental idea of general relativity. A friend of mine has objected that, from the purely mathematical point of view, we cannot infer from Einstein field equations (B.1) that rotation is a *fully* relative concept.

My answer is as follows. If we consider the field equations with the so-called cosmological term (where $\Lambda$ is a constant such that $|\Lambda| \leq 10^{-51}$ m$^{-2}$)

(C.1) $$G_{jk} + \Lambda\, g_{jk} = -\kappa\, T_{jk} \quad ,$$

when $T_{jk} \equiv 0$ the *only* solution which is *regular everywhere* is $g_{jk} \equiv 0$, and thus the principle of the relativity of inertia is perfectly satisfied [8].

**APPENDIX D**

Many learned works concern the research of *exact* undulatory solutions of Einstein field equations, see e.g. a paper by Bondi, Pirani, and Robinson [9]. Now, the characteristics of Einstein equations coincide with the characteristics of d'Alembert equation: accordingly, it is an obvious surmise to assume that, from the formal point of view, there exists a great variety of wavy solutions of the exact theory under the form of plane, cylindrical, spherical, etc. etc. waves. However, in order to attribute to these solutions a real *physical* meaning, it would be necessary to prove, first of all, that the chosen reference frames are physically privileged. Bondi, Pirani, and Robinson [9] seem to be aware of this fact, and actually they try to find a demonstration, but they only succeed in writing a sequence of sentences without logical strength: indeed, no such proof does exist.

There is a radical difference between the hypothetic gravitational waves and the electromagnetic waves, because the latter find their theoretical existence in the domain of *special relativity*, i.e. in the realm of the *symmetry* of Minkowski space-time.

### REFERENCES

[1] LEVI-CIVITA T., *Rend. Acc. Lincei*, (6) **11** (1930), 3 and 113; also in *Opere matematiche – Memorie e Note*, Volume 5° (Zanichelli, Bologna) 1970, p.77 and p.87.

[2] LOINGER A., http://xxx.lanl.gov/abs/astro-ph/9904207 (16 April 1999); IDEM, http://xxx.lanl.gov/abs/astro-ph/9810137 (8 Oct. 1998).

[3] BONNOR W.B. and SWAMINARAYAN N.S., *Z. Physik*, **177** (1964) 240.

[4] LORENTZ H. A., *Amst. Versl.*, **25** (1916) 468.

[5] LEVI-CIVITA T., *Rend. Acc. Lincei*, (5) **26** (1917) 381; also in *Opere matematiche – Memorie e Note*, Volume 4° (Zanichelli, Bologna) 1960, p.47.

[6] PAULI W., *Teoria della Relatività* (Boringhieri, Torino), 1958, p.262, and the literature quoted there.

[7] See the first paper cited in [2].

[8] PAULI W., *op. cit. in* [6], pp.274, 275.

[9] BONDI H., PIRANI F.A.E., and ROBINSON I., *Proc. Roy. Soc.*, **A 251** (1959) 519.

*————————————*